\documentstyle{mn}

\title[Magnetically modulated accretion in T Tauri stars]
  {Magnetically modulated accretion in T Tauri stars}
\author[C.J. Clarke et al.]
  {C.J. Clarke\thanks{Present address: School of Mathematical Sciences,
  Queen Mary \& Westfield College, Mile End Road, London, E1 4NS.},
  P.J. Armitage, K.W. Smith and J.E. Pringle\\
  Institute of Astronomy, Madingley Road,
  Cambridge CB3 0HA
  }
\date{Accepted 1994 October 26. Received 1994 August 18}
\pagerange{\pageref{firstpage}--\pageref{lastpage}}
\pubyear{1994}

\begin{document}

\label{firstpage}

\maketitle

\begin{abstract} We examine how accretion on to T Tauri stars may be
modulated by a  time-dependent `magnetic gate' where the inner edge of the
accretion disc is disrupted by a varying stellar field. We show that magnetic
field variations on time-scales $\ll 10^5$ yr can modulate the  accretion
flow, thus providing a possible mechanism both for the  marked photometric
variability of T Tauri stars and for the  possible conversion of T Tauri
stars between classical and weak line status. We thus suggest that archival
data relating to the spectrophotometric variability of T Tauri stars may
provide an indirect record of magnetic activity cycles in low-mass
pre-main-sequence stars.

\end{abstract}

\begin{keywords}

 stars: formation -- stars: T Tauri -- stars: rotation --
 stars: magnetic fields -- accretion discs
\end{keywords}

\section{Introduction}

The relationship between classical and weak line T Tauri stars (henceforth
CTTs and WTTs) remains an open question in the study of pre-main-sequence
stellar evolution. Whilst a number of spectral    diagnostics suggest that
CTTs are associated with material close to the stellar photosphere, it is not
all clear whether the weak line systems (which lack such signatures) are
necessarily devoid of circumstellar material at all radii and are thus,
permanently and irreversibly, truly naked systems. The strongest argument
that this is {\it not} the case is provided by the observation that a number
of T Tauri systems have been observed to switch between strong and weak line
status on a time-scale of decades \cite{hb}. Since this time-scale is
considerably less than even the fastest (dynamical) time-scale associated
with any reservoir of material at large radius it is clear that, in these
systems, the transition between CTT and WTT status is controlled by processes
close to the star. By implication, therefore, at least some stars classified
as WTTs must be indistinguishable  on the large scale from some CTTs. Further
support for this view is provided by the overlap in stellar parameters
between the WTTs and CTTs as a class, both in terms of their location on the
Hertzprung-Russell (HR) diagram and in terms of their angular momentum per
unit mass.  Furthermore the detection of spectral excesses in some WTTs at
submillimetre or near-infrared wavelengths has been claimed as evidence that
some systems are `weak' in terms of their circumstellar material close to the
star but are nevertheless associated with matter at large radii
\cite{strom,beckwith},  although this claim remains controversial
\cite{walter}.

In this paper we propose a simple mechanism that can allow systems to switch
between weak line and strong line status, and may also give rise to the
pronounced photometric variability of T  Tauri stars on a time-scale of
decades. We argue that variations in both the lines and continuum are related
to accretion processes close to the star, and that these are modulated by a
time-dependent magnetic `gate' at the inner edge of an accretion disc. In
this model the disc is disrupted by the magnetic field of the central star at
a distance of a few stellar radii; material can, however, only flow down the
field on to the star  if the star is rotating more slowly than the local
Keplerian  frequency of the disc at the magnetosphere. If, therefore, the
field  is variable, the changing location of the magnetosphere can give rise
to alternating phases of storage of material in the disc and of
accretion on to the star. We thus suggest that the variability pattern of T
Tauri stars may provide indirect evidence of the cycles of magnetic activity
in pre-main-sequence stars, analogous to the similar record inferred from
chromospheric variations of main-sequence dwarfs  \cite{baliunas88}.

In Section 2 we set out the model in more detail, estimating the field
strengths and variability time-scales for which the magnetic gate would work
in the way outlined above. In Section 3 we  discuss the model in relation to
a variety of observational parameters of T Tauri  stars.

\section{ The Model}

In this section we set out the model of magnetically coupled  T Tauri-disc
systems and derive the conditions under which such a model can modulate the
accretion flow on to the central star.

We assume that the central star possesses an ordered magnetic field on large
scales whose axis is aligned with the rotation axis of the star. Furthermore,
following Ghosh \& Lamb \shortcite{gl} and subsequent authors  we assume that
the field is able to penetrate the disc as a result of a variety of
micro-instabilities that ensure efficient mixing between the field and disc
plasma. Apart from a narrow region around corotation between disc and star,
the mismatched angular velocity between field lines (corotating with the
star) and Keplerian disc material  ensures that the field exerts a torque on
the disc, transferring  angular momentum from/to the disc respectively
inward/outward of corotation.

In such a situation, a disc of material flowing inwards under the action of
viscosity  is first significantly perturbed by the field at radius $r_{\rm
m}$ where the magnetic torque becomes comparable with the  local viscous
torque \cite{bepp,campbell}. (Note that, owing to the ordering of the
magnitude of the viscous torque per unit volume and  the thermal  and
dynamical energy densities in accretion discs, such a criterion is satisfied
when the magnetic pressure is negligible compared with either the thermal or
ram pressure in the disc; one can therefore use unperturbed disc structure
equations in the estimation of $r_{\rm m}$.) For steady-state disc flow
around a star of mass $M_*$,  at accretion rate $\dot M$, the differential
torque across an annulus of width ${\rm d}r$ is given by
\begin{equation}
{{{\rm d}G_{\nu}}\over{{\rm d}r}} ={{\dot M}\over{2}}
\biggl({{GM_*}\over{r}}\biggr)^{1/2},
\label{1}
\end{equation}
where $G_{\nu}$ denotes the torque due to viscous processes in the disc. The
corresponding differential magnetic torque is
\begin{equation}
{{{\rm d}G_{m}}\over{{\rm d}r}}=r^2 B_{\phi} B_z
\label{2}
\end{equation}
where $B_z$ and $B_{\phi}$ are the vertically averaged  poloidal and toroidal
components of the field respectively. Equating these quantities, the
magnetospheric radius then satisfies
\begin{equation}
B_zB_{\phi}|_{r_{\rm m}} =
 {\dot M\over{2}} \biggl({{GM_*}\over{r_{\rm m}^5}}\biggr)^{1/2}.
\label{3}
\end{equation}
We follow previous authors (e.g. Cameron \& Campbell
1993; K\"onigl 1991) in assuming that, whatever the chaotic field structure
near the stellar surface, the residual poloidal field at several stellar
radii is approximately dipolar:
\begin{equation} B_z|_{z=0} = B_*\biggl({{r_*}\over{r}}\biggr)^3
\label{4}
\end{equation}
(where $B_*$ is the
equatorial dipole field at the stellar surface, $r_*$) and that, moreover,
the equilibrium poloidal field threading the disc is close to its unperturbed
value \cite{gl,campbell}. Evaluation of the toroidal field, however, requires
several assumptions about the processes determining the equilibrium value of
$B_{\phi}$ in a shearing velocity field. Whereas the mechanism for the
generation of toroidal field is uncontroversial (i.e. differential rotation
between the footpoints of field lines anchored respectively on the star and
disc), a number of mechanisms have been suggested for  limiting the growth of
$B_{\phi}$, including reconnection of twisted field lines in the
magnetosphere \cite{lp}, magnetic buoyancy or turbulent magnetic diffusion in
the disc \cite{campbell,rlb} or else the action of a poorly specified
anomalous resistivity, whose role is to dissipate $B_{\phi}$ on a time-scale
comparable to the Alfv\'en crossing time of the disc height \cite{gl}. Here
we assume that magnetic reconnection in the magnetosphere is the dominant
limiting process, which implies $B_{\phi} \sim B_z$ since, away from
corotation, $B_{\phi}$ is both generated and destroyed on a dynamical
time-scale.  Thus equation (\ref{3}) becomes
\begin{equation} B_z^2|_{r_{\rm m}} =
 {\dot M\over {2}} \biggl({{GM_*}\over{r_{\rm m}^5}}\biggr)^{1/2}.
\label{5}
\end{equation}
It is immediately apparent that this expression is
within a numerical factor of order unity of that obtained if one calculates
the field strength required to dominate magnetically a spherical inflow of
the same $\dot M$ \cite{do}, thus justifying, to  order of magnitude, earlier
works that have applied the spherical Alfv\'en radius in the context of disc
accretion (e.g. K\"onigl 1991). A particularly convenient aspect of this
result is that $r_{\rm m}$ depends only on the field strength and $\dot M$:
\begin{equation}
r_{\rm m} = \biggl({{2B_*^2r_*^6}\over {(GM_*)^{1/2}\dot M}}\biggr)^{2/7},
\label{6}
\end{equation}
and is independent of the internal
structure of the disc giving rise to this steady-state accretion rate $\dot
M$.

We now consider the response of such a magnetically coupled star-disc system
to cyclical variations in the field strength of the central star. According
to equation (\ref{6}) such changes produce cyclical variations in the
magnetospheric radius $r_{\rm m}$.  At strong field phases of the cycle,
$r_{\rm m}$ may lie outside the radius of corotation between disc and star,
$r_{\Omega}$, so that the action of the field is then to impart angular
momentum to the disc. At such phases, therefore, material is held up in the
disc by the strong magnetic torques at its inner edge and is unable to flow
down the field on to the star.  As the field subsequently reduces, material
can again  flow inwards under the action of viscosity;  once inside
$r_{\Omega}$ the  magnetic torques change sign so that disc material  can be
decelerated and again flow on to the star. Thus modulation of the field can
give rise to an intermittent accretion pattern, with strong field phases in
which material is stored in the disc alternating with weak field phases in
which the stored material flows on to the star.

Several criteria have to be satisfied, however, in order for this process to
operate in the way outlined above. Since the modulation of accretion depends
on the alternate displacement of $r_{\rm m}$ inward and outward of corotation
it is necessary, for order-unity field variations to achieve this effect,
that $r_{\Omega}$ is close to the mean value of $r_{\rm m}$. Secondly, it is
also necessary that the field variations are sufficiently rapid for
variations of $r_{\rm m}$ to occur at approximately constant $r_{\Omega}$.
Below we consider each of these issues in the context of T Tauri stars.

For a star rotating with period $P_*$, the corotation radius is
\begin{equation}
r_{\Omega} = \biggl({{GM_*P_*^2}\over{4\pi^2}}\biggr)^{1/3}.
\label{7}
\end{equation}
Thus (comparing equations \ref{6} and \ref{7})  it
is evident that, for parameters  typical of classical T Tauri stars ($\dot M
\sim 10^{-7} M_*$ yr$^{-1}$, $P_* \sim 8$ d), $r_{\Omega}$ and $r_{\rm m}$
are of similar magnitude for surface field strengths of a few hundred gauss.
Such field strengths are within the upper limits derived from Zeeman effect
\cite{jp,bmv} and polarization measurements \cite{bl} in T Tauri stars, thus
giving support to the idea that T Tauri rotation is controlled by
magnetically coupled disc braking.

We now consider the maximum time-scale for field variations that    could
give rise to intermittent material flow on to the star: if the field were to
change sufficiently slowly then the star's rotation  could re-adjust so as to
maintain a fixed relation between $r_{\Omega}$ and $r_{\rm m}$ and hence
maintain a steady flow from disc to star. For the case $r_{\rm m} >
r_{\Omega}$ (for which there is no accretion on to the star)    the
time-scale for stellar spindown, $t_{\Omega}$, may be estimated as follows:
assuming that field-disc interaction is concentrated over a region of radial
extent $r_{\rm m}$ then  (from equations \ref{2} and \ref{5}) the spindown
torque exerted by the field is $\sim \dot M r_{\rm m}^2 \Omega_{\rm m}$,
where $\Omega_{\rm m}$ is the Keplerian frequency at the magnetosphere.  Thus
\begin{equation}
t_{\Omega} \sim {{fk^2M_*}\over{\dot M}}
\label{8}
\end{equation}
where $k\sim{0.4}$ \cite{heuvel} is the dimensionless radius
of gyration of the star and $f (<1)$ is the ratio of the specific angular
momentum of the stellar surface  to that of material in Keplerian rotation
at  the magnetosphere. (Note that, for variations on time-scales $\ll 10^6$
yr, the partial offset of this spindown by spinup due to the star's
contraction on its Kelvin-Helmholtz time-scale may be neglected.) For a T
Tauri star of rotation period $\sim 8$ d, $r_*/r_{\Omega} \sim 0.2 $ and
thus, for $r_{\rm m}$ modestly beyond $r_{\Omega}$,
\begin{equation} f =\biggl({{r_*}\over
{r_{\Omega}}}\biggr)^2\biggl({{r_{\Omega}} \over
{r_{\rm m}}}\biggr)^{1/2} \sim 0.04.
\label{9}
\end{equation}
Hence, for a T Tauri
disc with accretion rate $\sim 10^{-7} M_*$ yr$^{-1}$, $t_{\Omega} \sim 10^5$
yr. A similar argument may be made for the case  where $r_{\rm m} <
r_{\Omega}$, modified by the fact that in this case material flows from the
disc on to the star. This introduces an additional angular momentum input
rate of $\dot M r_{\rm m}^2 \Omega_*$ (which is, however, smaller than the
spinup torque from the field by a factor  $\Omega_*/{\Omega_{\rm m}}<1$) plus
an additional spinup torque due to the star's (adiabatic) contraction in
response to accretion; this latter effect occurs on a time-scale $\sim
M_*/\dot M \sim 10^7$ yr. The dominant contribution to changing the stellar
period is thus the magnetic torque (whether $r_{\rm m}$ is greater or less
than $r_{\Omega}$) and the appropriate time-scale is $\sim 10^5$ yr in either
case. We therefore deduce that field variations on time-scales substantially
less than $10^5$ yr occur at approximately constant rotation, and thus result
in the modulation of the relative values of $r_{\rm m}$ and $r_{\Omega}$.

\section{Discussion}

In the previous section we showed that, for field strengths of a few hundred
gauss, order-unity variations of the field on  time-scales  $\ll 10^5$ yr
would strongly modulate the accretion flow on to  the central star. Further
characterization of the time-dependent flow and associated photometric
variability requires detailed hydrodynamic modeling (Armitage et al., in
prep). The main implications of this conclusion are, first, that
spectrophotometric variations of T Tauri stars may thus provide a record of
the magnetic activity cycle of T Tauri stars and, secondly, that at least
some weak line systems  are interconvertible with strong line systems. We
examine each of these issues below.

The temporal behaviour of magnetic fields  in T Tauri stars is not known,
apart from the rapid variations deduced (by analogy with solar
chromospheric/coronal flares) from short-time-scale ultraviolet \cite{wska}
and X-ray (Montmerle et al. 1983, 1993) flares in these objects. Further
analogy with magnetic activity in main-sequence stars leads one to anticipate
that T Tauri magnetic fields would also be variable on a range of longer
time-scales, as witnessed by the chromospheric variations of late-type dwarfs
(time-scales $\sim 10$ yr: Baliunas 1988) and the history of intermittent
activity on longer time-scales contained in the sunspot record. The
qualitative  similarity between the convective dynamos believed to be
responsible for magnetic fields in both main-sequence and pre-main-sequence
low-mass stars leads one to anticipate a similarly variable pattern in T
Tauri stars.

Pronounced variability, on a range of time-scales, is one of the defining
characteristics of T Tauri stars as a class. Apart from the rapid flares
alluded to above, which may be interpreted as localized surface phenomena on
the star, it is hard to avoid the conclusion that this variability is related
to variable accretion on to the  star since, in the absence of accretion, it
is difficult to envisage  what processes in the stellar interior could
account for such  large-amplitude variations in the bolometric output of the
star on these time-scales. Variable accretion can result either from
magnetic variations (as described here) or else from variable  flow in the
accretion disc, although it is not at all obvious  why the accretion flow
should be intrinsically  intermittent on these time-scales. We note that the
hypothesis that T Tauri variability reflects the  stars' magnetic activity
opens up the possibility of using the photometric variability record of T
Tauri stars as    a diagnostic of magnetic activity cycles in T Tauri stars,
in much the same way as chromospheric variations  are used in main-sequence
stars. However, whereas the latter record only extends back a few decades,
and for a limited sample of stars that have been deliberately monitored for
this purpose, published photometry for T Tauri stars, together with archival
plate material, means that a database extending back over a century is
available for a large sample of stars.

We next turn to the question of the interconvertibility of  strong and weak
line T Tauri stars.  The strongest evidence for the possibility of switching
between the two states is provided by the half a dozen systems in the Herbig
\& Bell \shortcite{hb} catalogue, for which it is noted that their current
designation as weak or strong line does not accord with the results of
objective prism surveys in previous decades. Since these remarks do not
relate to systematic monitoring programmes they cannot be used to derive duty
cycles, but just illustrate that such transitions can occur on time-scales of
decades. Further plausibility arguments about the possible interconversion of
WTTs and CTTs  depend on the overlap in properties (apart from those
indicative of material flow at small radii) between  the two classes of
object. It is well known, for example, that CTTs and WTTs co-exist in the
same region of the HR diagram (apart from a  WTT-only region near the
main-sequence), thus arguing against a simple picture in which a universal
clock controls an  irreversible transition  of stars from WTT to CTT status.
The rotational properties of CTTs and WTTs  also overlap: whereas the WTTs
rotate faster as a class \cite{bouvier,edwards}, about half of them have
angular momenta  that overlap with the CTTs. Since the slow  rotation of T
Tauri stars is commonly attributed to a history in which they have been
braked by magnetic coupling to associated discs \cite{konigl,cc}, the low
angular momentum WTTs have probably undergone a similar history of disc
braking. Such data suggest a picture in which objects classified as WTTs are
a mixture of genuinely naked systems (devoid of material at all radii) and
those that, retaining a reservoir of circumstellar material, are capable of
being resuscitated as strong line systems.

The above picture suggests some correlations between the diagnostics of
material at small radii and the rotational properties of T Tauri stars. For
example, if the low angular momentum T Tauri  stars have evolved to a state
of rotational quasi-equilibrium (i.e.  if the time-averaged magnetospheric
radius is close to corotation, so that the the spinup and spindown torques
exerted by disc material nearly cancel) then the magnetic field should
disrupt the disc out to  larger radii  in more slowly rotating systems, thus
increasing the volume of the region of quasi-spherical infall. Since ${\rm H}
\alpha$  emission is believed to arise from material flowing nearly radially
near the star (e.g. Calvet \& Hartmann 1992) one would therefore anticipate a
larger ${\rm H} \alpha$ flux in slowly rotating systems of given accretion
rate, since the solid angle intercepted by inflowing material is larger than
in more rapidly rotating systems. On the other hand, the range of rotation
periods among CTTs may indicate that these systems have not all evolved to a
state in which the magnetosphere lies at corotation; in this case the
time-averaged magnetosphere would lie outside corotation in more  rapidly
rotating systems. Since (see Section 2) accretion on to the star does not
occur  when the magnetosphere is  outside corotation it follows that, in a
magnetic cycle, the fraction of the cycle for which accretion can occur is
lower for rapidly rotating systems.  Thus in this case, also, one would
expect weaker diagnostics of accretion close to the star for more rapidly
rotating stars. It is notable that these predictions are borne out by the
observed decline in the ratio of ${\rm H} \alpha$ (indicative of material
close to the star) to excess $K$ magnitude (indicative of the reservoir of
material at larger radii) with increasing stellar rotation \cite{bouvier}.

Finally we note that such a model is not unique in predicting a variety of
spectral energy distributions according to the inner radius to  which the
disc extends. The possibility of systems that are `weak' on small scales and
`strong' on large scales   has been recognized by various authors  who have
sought to identify  spectral `gaps' in some T Tauri stars as clearing of the
inner disc by planetary formation (e.g. Marsh \& Mahoney 1992). Such a
picture does not, however, unlike the variable magnetic disc model described
here, allow the reversible exhaustion/replenishment of material on small
scales as suggested by the variability data described above.

\label{lastpage}

\end{document}